\documentclass[prb,twocolumn]{revtex4-1}
\pdfoutput=1
\usepackage{float}
\usepackage{graphicx}
\usepackage{epstopdf}
\usepackage{amsmath}
\usepackage{mathtools}
\usepackage{subfigure}
\usepackage[colorlinks,linkcolor=blue,anchorcolor=blue,citecolor=blue,urlcolor=blue]{hyperref}

\renewcommand{\vec}[1]{{\boldsymbol{#1}}}
\newcommand{\bk}{\vec{k}}
\newcommand{\br}{\vec{r}}
\newcommand{\heading}[1]{\paragraph*{#1}}

\def\be{\begin{equation}}
\def\ee{\end{equation}}
\def\bea{\begin{eqnarray}}
\def\eea{\end{eqnarray}}

\def\nn{\nonumber}

\begin{document}
\title{The fate and heir of Majorana zero modes in a quantum wire array}
\author{Da Wang}
\affiliation{Department of Physics, University of California, 
San Diego, CA92093}
\author{Zhoushen Huang}
\affiliation{Department of Physics, University of California, 
San Diego, CA92093}
\author{Congjun Wu}
\affiliation{Department of Physics, University of California, 
San Diego, CA92093}

\begin{abstract}
Experimental signatures of Majorana zero modes in a single superconducting
quantum wire with spin-orbit coupling have been reported as zero bias 
peaks in the tunneling spectroscopy. 
We study whether these zero modes can persist in an array of 
coupled wires, and if not, what their remnant could be. 
The bulk exhibits topologically distinct gapped phases and an 
intervening gapless phase.
Even though the bulk pairing structure is topological, 
the interaction between Majorana zero modes and superfluid
phases leads to spontaneous time-reversal symmetry breaking.
Consequently, edge supercurrent loops emerge and edge Majorana fermions 
are in general gapped out except when the number of chains is odd, 
in which case one Majorana fermion survives.
\end{abstract}
\maketitle

\heading{Introduction}%
Majorana fermions are intriguing objects because they are their own 
antiparticles. 
In condensed matter physics, Majorana fermions arise not as 
elementary particles, but rather as superpositions of electrons 
and holes forming the zero mode states in topological 
superconducting states.
Majorana fermions were first proposed to exist in vortex cores and 
on boundaries of the $p$-wave Cooper pairing systems \cite{Volovik1999,Read2000,Kitaev2001}.
More recently, they were also predicted
in conventional superconductors in the presence of strong spin-orbit 
(SO) coupling and the Zeeman field
 \cite{Fu2008,Lutchyn2010,Sau2010,Oreg2010,Alicea2010,Mao2012}.
In cold atom physics, SO coupling has been realized by using atom-laser
coupling \cite{Lin2009,Lin2011,Wang2012,Cheuk2012,Goldman2013}.
This progress offers an opportunity to realize and manipulate Majorana
fermions in a highly controllable manner, which has attracted
a great deal of attention both theoretical and experimental
\cite{Zhang2008,Sato2009,Zhu2011,Jiang2011,Gong2012,Seo2012,
Wei2012,Liu2012,Liu2012b,Romito2012,Kraus2013,Li2013,Mizushima2013,Qu2013}.

Experimental signatures of Majorana zero modes have been reported as 
zero bias peaks in the tunneling spectroscopy of a single quantum 
wire with strong SO coupling which is either coupled with 
an $s$-wave superconductor through the proximity effect 
\cite{Mourik2012,Deng2012,Das2012,Rokhinson2012,Finck2013,Lee2012,
Lee2014}, or, is superconducting by itself \cite{Rodrigo2012,*Rodrigo2013}.
A further study of an array of quantum wires is natural
\cite{Li2013,Mizushima2013,Seroussi2014,Lutchyn2011,Stanescu2011,
Tewari2012,Diez2012,Qu2013,Kells2013},
in particular for the purpose of studying
interaction effects among edge Majorana zero 
modes \cite{Fidkowski2010,Ryu2012,Qi2013,Li2013}.
Topological states in an array of parallel wires in 
magnetic fields in the fractional quantum Hall regime have been 
studied recently \cite{klinovaja2013,klinovaja2013a}.
Without imposing self-consistency, flat bands of Majorana zero edge 
modes have been found for the uniform pairing
as well as the Fulde-Ferrell-Larkin-Ovchinnikov
pairing \cite{Lutchyn2011,Stanescu2011,Tewari2012,Diez2012,Qu2013},
because under time-reversal (TR) symmetry these Majorana zero modes
do not couple.

However, the band flatness of the edge Majorana zero modes
is unstable due to interaction effects.
Li and two of the authors proposed the mechanism of spontaneous
TR symmetry breaking for the gap opening in the edge Majorana flat
bands\cite{Li2013}.
Even in the simplest case of spinless fermions 
without any other interaction channels, the coupling between Majorana 
zero modes and the pairing phase spontaneously generates staggered 
circulating currents near the edge such that Majorana modes can 
couple to each other to open the gap due to the breaking of TR symmetry. 
Similar results are also obtained recently in Refs. 
[\onlinecite{Mizushima2013,Potter2013}].
The mechanism of gap opening based on spontaneous TR symmetry
breaking  also occurs in the helical edge modes of 
the 2D topological insulators under strong repulsive interactions 
which leads to edge magnetism
\cite{Wu2006,Xu2006}.

\heading{Main results} 
In this article, we investigate a coupled array of $s$-wave superconducting 
chains with intra-chain SO coupling and external Zeeman field. 
We consider both the proximity-induced superconductivity and the intrinsic one. 
For the proximity-induced case, the array is placed on top of a bulk 
superconductor, the phase coherence induces a  nearly 
uniform pairing distribution in the quantum chains, $\Delta_\br = \Delta$. 
The bulk band structure exhibits several topologically 
distinct gapped phases intervened by a gapless phase. 
In the gapless phase, edge Majorana zero modes interpolate between 
nodes in the bulk energy spectrum. 
In the topological gapped phase, they extend into a flat band
across the entire edge Brillouin zone.
On the other hand, if either the phase coherence of the bulk
superconductor is weak, or the superconductivity is intrinsic, such as 
in the case of Pb nanowires \cite{Rodrigo2012,*Rodrigo2013}
or cold atom systems near Feshbach resonance \cite{Jiang2011},
then $\Delta_\br$ has to be solved self-consistently. 
We find that when the bulk is in the topological gapped phase,
the phase distribution of pairing order parameters is inhomogenous
along the edge exhibiting TR symmetry breaking. 
It induces edge currents and gaps out the edge Majorana zero modes except 
when the chain number is odd, in which case 
one Majorana zero mode survives. 
If the bulk is in the gapless phase, in general TR breaking is also observed
but not always, because Majorana modes associated with opposite winding 
numbers can coexist on the same edge which can be coupled by TR invaraint 
perturbations.

\begin{figure}
\includegraphics[width=\linewidth,height=0.7\linewidth]{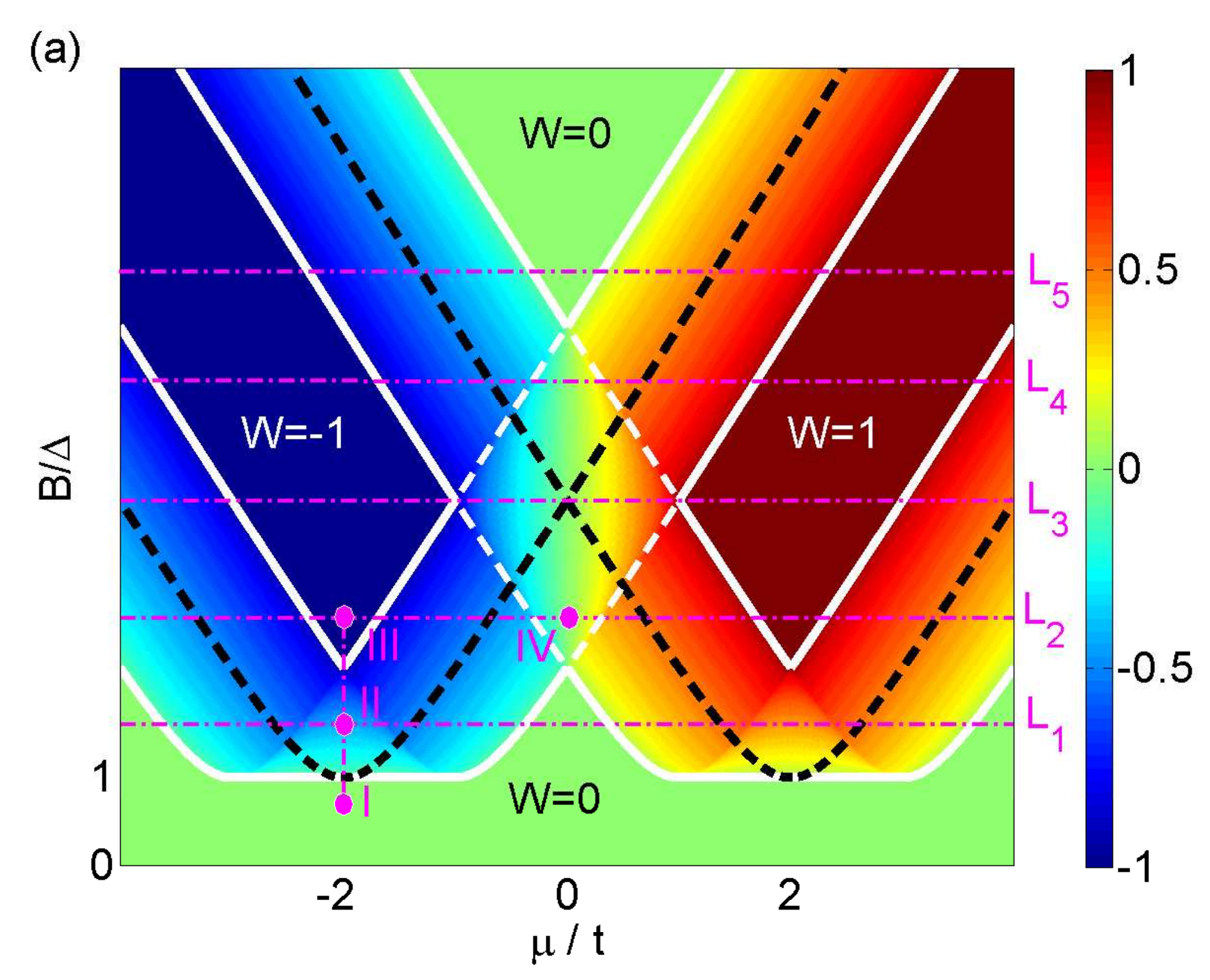}
\includegraphics[width=\linewidth,height=0.7\linewidth]{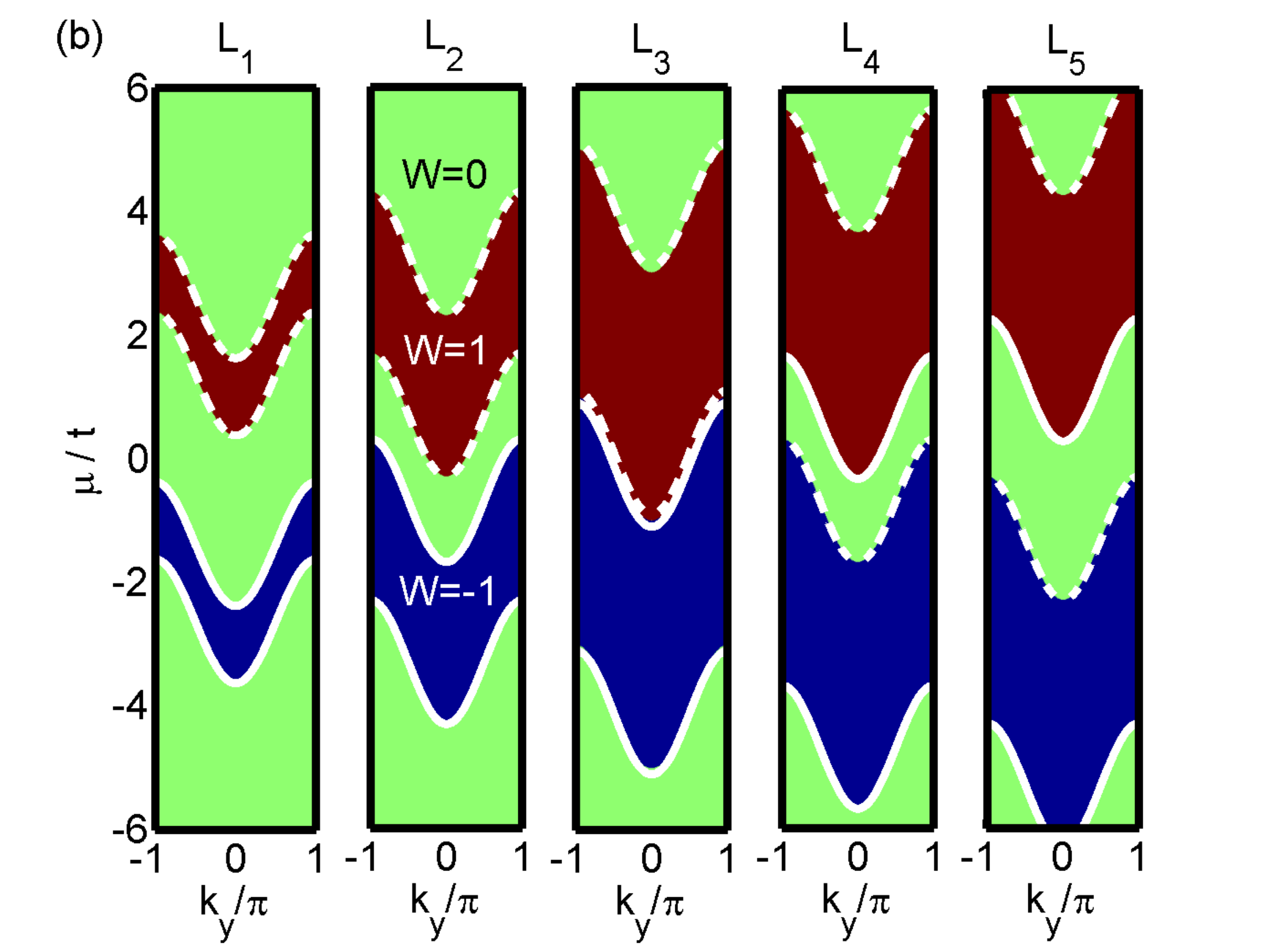}
\caption{
($a$) Bulk phase diagram of the 2D Hamiltonian Eqs. \ref{eq:ham_band}
and \ref{eq:ham_ex} in the $\mu$-$B$ plane with $B>0$, 
and that with $B<0$ is symmetric with respect to the axis of $B=0$.
The parameter values are $t=1$, $t_\perp=0.5$, $\lambda=2$, $\Delta=0.5$.
(a) The white solid lines enclose the gapless phase and separate the rest 
into two topologically trivial gapped phases and two non-trivial
phases, respectively.
Inside the gapless phase, states in the diamond enclosed by the white 
dashed lines exhibit edge modes associated with opposite winding numbers,
and those outside the diamond only exhibit edge modes associated
with same winding number.
Color scale encodes the momentum averaged winding number $r$ defined 
in Eq. \ref{eq:topo_avg}.
The gapless phase is suppressed as decreasing $t_\perp$,
and it is compressed into the black dashed line at $t_\perp=0$
(the single chain limit).
Points I $\sim$ IV are used in Fig. \ref{fig:ek_noscf}.
(b) $W_{k_y}$ v.s. $k_y$ and $\mu$ are shown along
the lines of $L_1\sim L_5$ in ($a$), respectively.  
The white solid and dashed boundaries of the regions of 
$W_{k_y}=\pm1$ represent that the gap closing points are
located at $(0,k_y)$ and $(\pi,k_y)$, respectively.
\label{fig:bulkphasediagram}}
\end{figure}

\heading{Model of quantum wire array}%
Consider an array of SO coupled chains with the proximity effect
induced $s$-wave pairing along the $x$-direction, which are juxtaposed 
along the $y$ direction. 
The band Hamiltonian is
\bea
H_0&=&-\sum_{\br \sigma} t \left(c_{\br \sigma}^\dag c_{\br +\hat{x},\sigma} 
+ h.c.\right) - \mu c_{\br \sigma}^\dag c_{\br \sigma} \nonumber \\
&-&\sum_{\br } i\lambda \left(c_{\br \uparrow}^\dag c_{\br +\hat{x},\uparrow} 
- c_{\br \downarrow}^\dag c_{\br +\hat{x},\downarrow}\right) + h.c.
\nn \\
&-&\sum_{\br \sigma} t_\perp \left( c_{\br \sigma}^\dag c_{\br +\hat{y},\sigma} 
+ h.c. \right),
\label{eq:ham_band}
\eea
where $\vec r$ is the lattice site index; $\sigma=\uparrow\,,\,\downarrow$ 
labels two spin states;  $t$ and $t_{\perp}$ are intra- and 
inter-chain nearest neighbor hoppings, respectively,
and $\mu$ is the chemical potential. 
$\lambda$ here is the SO coupling, which we choose to lie only in the $x$-direction. 
This uni-directional SO coupling is a natural
setup in cold atom experiments using atom-laser interaction.\cite{Jiang2011,Qu2013}
The external field part of the Hamiltonian is
\bea
H_{ex}=\sum_{\br}  
\Delta_\br( c_{\br \uparrow}^\dag c_{\br \downarrow}^\dag + 
h.c.)-B (c_{\br \uparrow}^\dag c_{\br \downarrow} 
+ h.c.), 
\label{eq:ham_ex}
\eea 
The first term accounts for superconducting pairing,
  where $\Delta_{\vec r}$ is the $s$-wave pairing on site $\br$, and
  can be induced either through proximity effect or intrinsically. For
  the proximity induced superconductivity, we take $\Delta_\br$ to be
  spatially uniform, which is a commonly used approximation. For
  intrinsic superconductivity, $\Delta_\br$ will be solved
  self-consistently. The second term arises from an external Zeeman
  field $B$, which can also be simulated using
  atom-laser coupling\cite{Jiang2011}. 

\heading{Uniform pairing}%
Let us first consider a uniform pairing $\Delta_\br=\Delta$
which can be chosen as real without loss of generality.
Under periodic boundary conditions in both $x$ and $y$-directions, 
the Hamiltonian Eqs.~\ref{eq:ham_band} and \ref{eq:ham_ex} 
can be written in momentum space,
\bea
H=H_{band}+H_{ex}=\sum_{\bk}\psi_\bk^\dag h_\bk \psi_\bk, 
\eea
where
$\psi_\bk=[c_{\bk\uparrow},c_{\bk\downarrow},c_{-\bk\uparrow}^\dag,
c_{-\bk\downarrow}^\dag]^t$,
and
\begin{eqnarray}
h_\bk=T_\bk\tau_3+\Lambda_\bk\sigma_3-B\sigma_1\tau_3+\Delta\sigma_2\tau_2\ ,
\label{eq:hk}
\end{eqnarray}
The two sets of Pauli matrices $\sigma_i$ and $\tau_i$ ($i=1,2,3$) 
act in the spin and particle-hole spaces, respectively. 
$T_\bk$ and $\Lambda_\bk$ are given by 
\bea
T_\bk=-2t\cos k_x-2t_\perp\cos k_y-\mu ,
\eea
and 
\bea
\Lambda_\bk=2\lambda\sin k_x .
\eea

The energy spectrum of Eq. \ref{eq:hk} is
\bea
E_\bk^2 = T_\bk^2+\Lambda_\bk^2+B^2+\Delta^2  
\pm 2\sqrt{T_\bk^2\Lambda_\bk^2+T_\bk^2B^2+B^2\Delta^2}. \nn \\
\eea

Although $h_{k_x,k_y}$ does not carry 2D topological indices,
nevertheless, we consider the 1D index of $h_{k_x,k_y}$ at each fixed
value $k_y$. It is invariant under both particle-hole ($\Xi$) and TR
($\Theta$) symmetries: define 
\bea
\Xi = \tau_1,\quad \Theta = \sigma_1\tau_3, \label{eq:pseudoTR}
\eea
then 
\bea
\Xi h_{k_x,k_y}\Xi^{-1}=-\Theta h_{k_x,k_y}
\Theta^{-1}=-h_{-k_x, k_y}^*. 
\eea
Here both transformations satisfy
$\Theta^2=\Xi^2=1$. 
We should emphasize here that $\Theta$ is \emph{not} the
  physical time reversal, which should square to $-1$ for fermions
  with a half-integer spin. Here $\Theta$ is called ``time reversal''
  because it represents a symmetry operation which is anti-unitary and
  relates $\bk$ to $-\bk$.
$\Theta$ and $\Xi$ 
can be combined into a chiral symmetry 
defined as 
\bea
\mathcal{C}=\Xi\Theta, 
\eea
which gives 
\bea
\mathcal{C}h_{k_x,k_y}\mathcal{C}^{-1}=-h_{k_x,k_y} .
\eea 
These symmetries put $h_{k_x,k_y}$ at fixed $k_y$ in the BDI class as pointed
out in Ref. [\onlinecite{Tewari2012}], which is 
characterized by a $k_y$-dependent 1D topological index denoted as $W_{k_y}$. 
A unitary transformation is performed as
\bea
U=\mbox{e}^{i(\pi/4)\sigma_2}\,u\,\mbox{e}^{-i(\pi/4)\tau_1},
\eea 
where 
\bea
u=\frac{1}{2}(\sigma_0+\sigma_3)+\frac{1}{2}\tau_3(\sigma_0-\sigma_3).
\eea
It transforms $h_\bk$ into an off-diagonal form
\begin{eqnarray}
U^{-1}h_\bk U=\begin{bmatrix}
0&A_\bk \\A_\bk^\dag &0
\end{bmatrix}\ ,
\label{eq:hkoff}
\end{eqnarray}
where
\bea
A_\bk=\Delta\sigma_1-i(T_\bk\sigma_0+\Lambda_\bk\sigma_1+B\sigma_3).
\eea
$W(k_y)$ is defined as the winding
number of $\det A_{\bk}$ in the complex plane as $k_x$ sweeps a
$2\pi$ cycle, \emph{viz.},
\cite{Sato2011,Tewari2012}
\bea
W_{k_y}&=&-\frac{i}{2\pi}\int\limits_{\mathclap{k_x=0}}^{2\pi}\frac{\mbox{d}z(\vec{k})}{z(\vec{k})}\nn \\
&=&\frac{1}{2}\bigl[\mbox{sgn}(M_+)-\mbox{sgn}(M_-)\bigr]
\mbox{sgn}(\lambda \Delta)\ ,\quad
\label{eq:w-def}
\eea
where $z(k)=\det A_k/|\det A_k|$, in which 
\bea
\det A_k=B^2-T_k^2-(\Delta-i\Lambda_k)^2,
\eea
$M_\pm(k_y)$ are related to $\det A_{k_x,k_y}$ as
\bea
M_+(k_y)=\det A_{k_x=0,k_y},  \\M_-(k_y)=\det A_{k_x=\pi,k_y}.
\eea
$W_{k_y}=\pm 1$ requires the condition of $M_+(k_y)M_-(k_y)<0$, 
and then $h(k_y)$ is topologically nontrivial. 
$W_{k_y}$ changes discretely if a gap closing state
appears on the line of $k_y$ such that 
$M_+(k_y)=0$, or, $M_-(k_y)=0$.
The momenta of these states $(k_x,k_y)$ satisfy
that $k_x=0$ or $\pi$, and another condition 
$T_\bk^2+\Delta^2-B^2=0$ which determines $k_y$.

Based on $W_{k_y}$'s behavior over the range of $[-\pi,\pi)$, we 
plot the bulk phase diagram for the 2D Hamiltonian
Eqs. \ref{eq:ham_band} and \ref{eq:ham_ex} in the parameter plane
$\mu$-$B$ shown in Fig.~\ref{fig:bulkphasediagram}($a$).
The gapped phases are characterized by $k_y$-independent values of 
$W$: two phases with $W=\pm 1$ are weak topological pairing states,
and the other two with $W=0$ are trivial pairing states.
For the gapless phase, a momentum averaged topological number is defined as
\bea
r = \int\frac{\mathrm{d}k_y}{2\pi} W_{k_y}.
\label{eq:topo_avg}
\eea
The values of $W_{k_y}$ {\it v.s.} $\mu$ and $k_y$ are depicted 
in Fig.~\ref{fig:bulkphasediagram}($b$) along the line 
cuts $L_1\sim L_5$ in Fig. ~\ref{fig:bulkphasediagram} (a).
Usually, $W_{k_y}$ only changes the value by 1 at one step
as varying $k_y$, but
along line $L_3$, $W_{k_y}$ can directly change between 1 and -1
without passing $0$, which means two Dirac points $(0,k_y)$
and $(\pi,k_y)$ appear at the same value of $k_y$. 
Note that the SO coupling $\lambda$ is related to $W_{k_y}$ only through its sign (\textit{c.f.} Eq.~\ref{eq:w-def}), 
therefore the phase diagram Fig.~\ref{fig:bulkphasediagram} is independent of  
$\lambda$ (up to an overall sign flip). 
\footnote{The value of $\lambda$ determines the 
bulk gap in topological superconducting regime.
Therefore, for a given temperature $T$, $\lambda\gg T$
is required to observe this topological superconductivity physics.}

\begin{figure}
\includegraphics[width=\linewidth,trim=0 35 0 45]{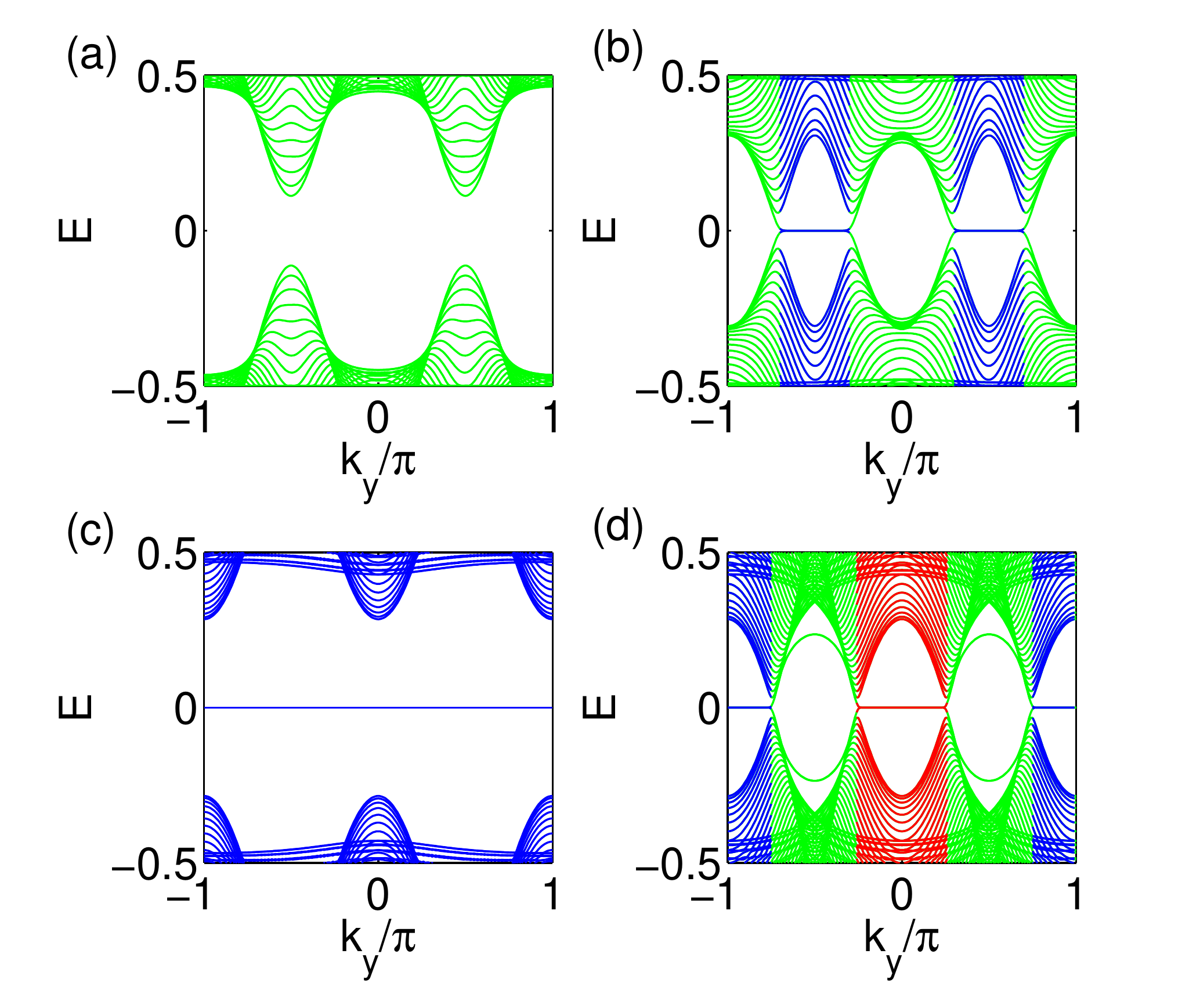} 
\caption{Edge spectra with the open and periodical boundary conditions 
along the $x$ and $y$-directions, respectively.  
($a$), ($b$), ($c$), and ($d$) correspond to points I, II, III, and IV
marked in Fig. \ref{fig:bulkphasediagram} ($a$), respectively. 
($a$) gapped trivial phase;
($b$) gapless phase with edge modes associated with the same winding number;
($c$) gapped weak topological phase;
($d$) gapless phase with edge modes associated with opposite winding numbers.
Parameters used are the same as those in Fig. \ref{fig:bulkphasediagram} $(a)$.
\label{fig:ek_noscf}}
\end{figure} 

Next we discuss edge spectra in the above different phases.
The open boundary condition is applied along the $x$-direction.
In the topological trivial phase shown in Fig. \ref{fig:ek_noscf}
($a$), the zero energy edge modes are absent, while they appear and 
run across the entire 1D edge Brillouin zone in the gapped weak 
topological pairing phase shown in Fig. \ref{fig:ek_noscf} ($b$).
In the gapless phase, flat Majorana edge modes appear in the regimes
with $W(k_y)=\pm 1$ and terminate at the gap closing points.
\cite{Sato2011,Sau2012,Yuan2014} 
These flat Majorana edge modes are lower dimensional Majorana
analogues of Fermi arcs in 3D Weyl semi-metals \cite{Wan2011,*Balents2011}. 
This analogy goes further as in both cases: the gapless phase 
intervenes topologically distinct gapped phases. 

The flat edge Majorana modes in the gapless phase can behave differently. 
In Fig. \ref{fig:ek_noscf} ($b$), all the edge flat Majorana 
modes are associated with the same value of $W_{k_y}$.
In this case, these Majorana modes on the same edge are robust 
at the zero energy if TR symmetry is preserved, which means that
they do not couple.
Nevertheless, TR symmetry may be spontaneously broken to gap
out these zero modes \cite{Li2013}.
On the other hand, for states inside the white dashed diamond
in Fig. \ref{fig:bulkphasediagram} ($a$), edge Majorana 
modes appear with both possibilities of $W_{k_y}=\pm 1$. 
In particular, in the case of $\mu=0$, the relation 
$W(k_y)=-W(k_y+\pi)$ holds for edge Majorana modes  
as shown in Fig. \ref{fig:ek_noscf} ($d$).
Majorana modes with opposite winding numbers on the same edge
can couple to each other even without TR breaking, 
and thus are not topologically stable. 
$r$ represents the net density of states 
of zero modes in the edge Brillouin zone which are stable
under TR-conserved perturbations.

\begin{figure}
\includegraphics[width=0.5\textwidth,trim=0 50 0 30,clip]{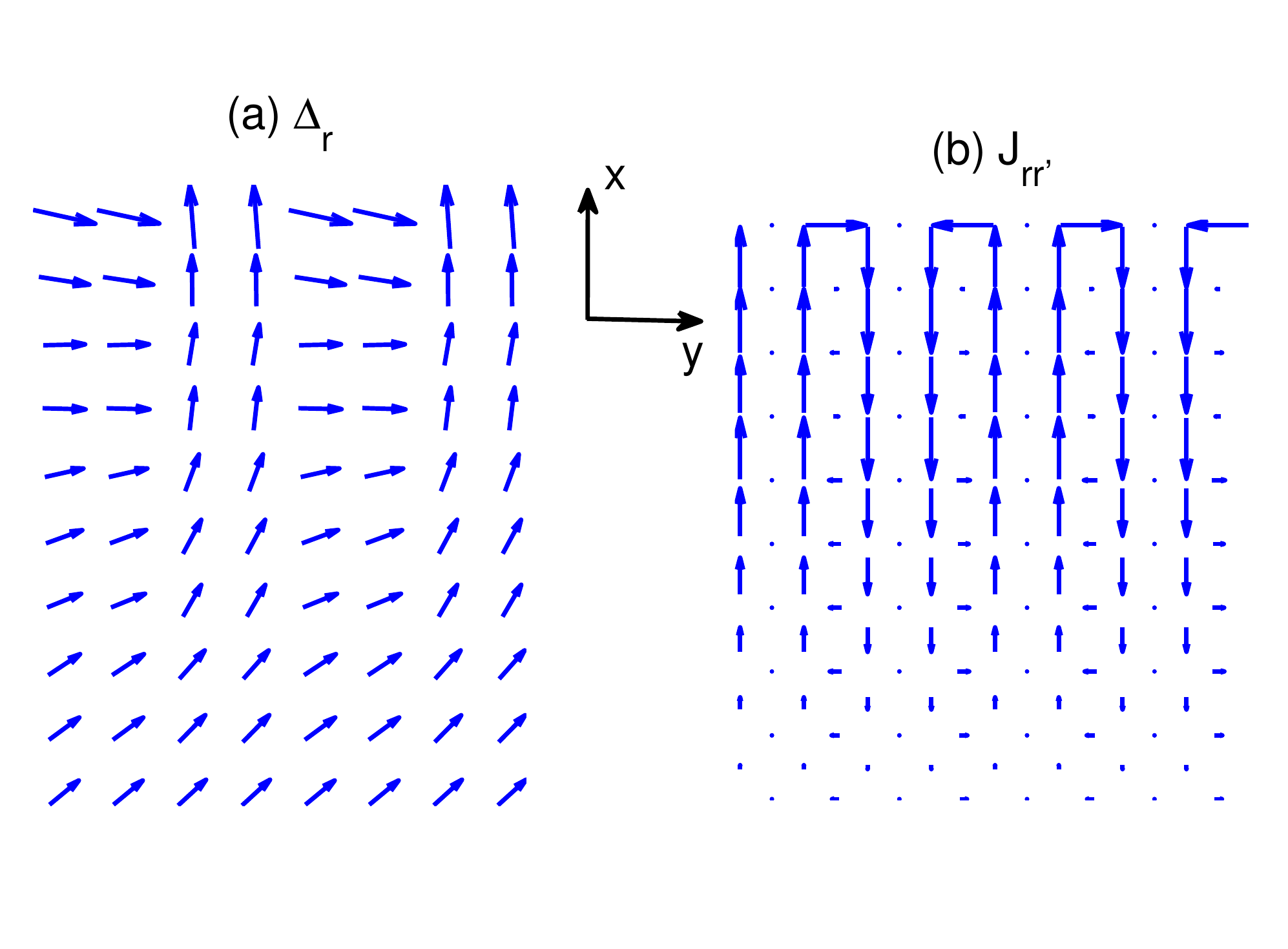}
\caption{Self-consistent solutions for $\Delta_{\br}$
($a$) and supercurrent $J_{\br\br^\prime} $ ($b$).
Parameter values are $L_x=120$, $L_y=8$, $B=1.25$, $t=1$, 
$t_\perp=0.5$, $\mu=-2$, $\lambda=2$, $g=5$.
Open and periodical boundary conditions are used along the $x$-direction
(vertical) and $y$-direction (horizontal), respectively. 
Only the first 10 sites from the upper edge are plotted.
The distributions of $\Delta_\br$ and $J_{\br\br^\prime}$ are reflection
symmetric with respect to the center line of the system.
($a$) Direction and length of each arrow represent the
phase and amplitude of $\Delta_\br$ on site $\br$.
Its distribution is nearly uniform in the bulk but exhibits 
spatial variations near the edge. 
($b$) Each arrow represents $J_{\br\br^\prime}$ on bond $\br\br^\prime$, 
which is prominent near the edge but vanishes in the bulk.
\label{fig:L8}}
\end{figure}

\heading{Self-consistent solution}
We now impose self-consistency on the pairing order parameter 
$\Delta_{\br}$, which is necessary for the case of intrinsic pairings. 
The pairing interaction is modeled as 
\bea
H_\Delta=-g\sum_{\br} n_{\br,\uparrow}n_{\br,\downarrow}, 
\eea
and the self-consistent 
equation is 
\bea
\Delta_\br=-g\langle G| c_{\br \downarrow}c_{\br \uparrow}|G \rangle,
\eea
where $\langle G|... |G\rangle$ means the ground state average. 
We have verified numerically that $\Delta_\br$ is nearly uniform inside 
the bulk.
Thus the bulk shares a similar phase diagram to the case of 
uniform pairing (\emph{cf.~}Fig.~\ref{fig:bulkphasediagram}), 
except that the values of $\Delta$ should be self-consistently determined.

Nevertheless, near edges $\Delta_{\br}$ varies spatially in the
self-consistent solutions.
If the bulk is in the topological gapped phase, the edge Majorana
zero modes can couple with each other by breaking TR symmetry
spontaneously as shown in Ref. [\onlinecite{Li2013}].
Because of the band flatness, this effect is non-perturbative. 
This will gap out the zero Majorana modes and lower the edge energy. 
The system converges to an inhomogeneous distribution of 
$\arg[\Delta_{\br}]$ near the edges as shown in 
Fig.~\ref{fig:L8} ($a$), even if this costs energy 
by disturbing the Cooper pairing \cite{Li2013}.
This edge inhomogeneity in the pairing phase leads to an 
emergent current pattern as depicted in Fig.~\ref{fig:L8} ($b$).

\begin{figure}[h!]
\includegraphics[width=0.8\linewidth]{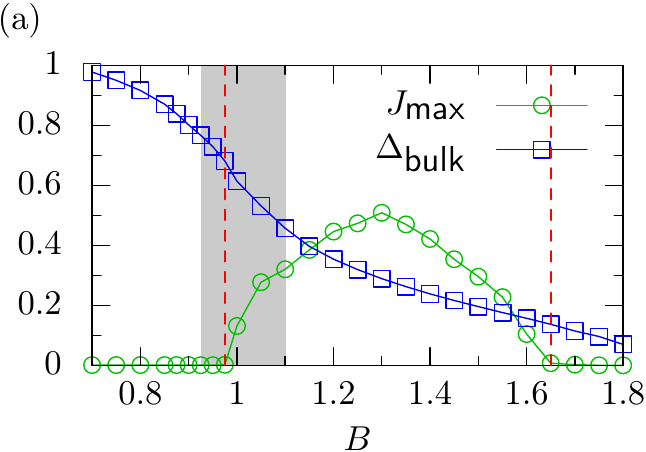}
\includegraphics[width=0.8\linewidth]{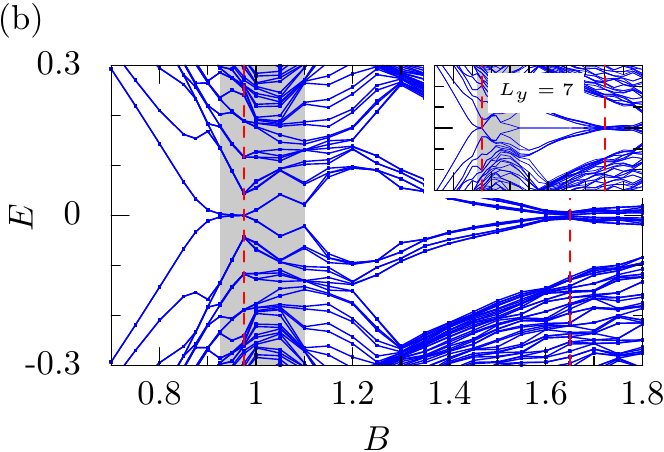}
\caption{Self-consistent solutions for coupled chains with varying
$B$-field.
Open and periodical boundary conditions are used along the $x$ and 
$y$-directions, respectively.
Parameters are $L_x=120$, $L_y=8$, $t=1$, $t_\perp=0.5$, $\mu=-2$, 
$\lambda=2$, and $g=5$.
In both ($a$) and ($b$), the bulk gapless phase is marked as the shaded 
region, which separates the topologically trivial (on its left) and 
nontrivial (on its right) gapped phases. 
($a$) The bulk pairing $\Delta_{bulk}$ and the
characteristic edge current magnitude $J_{max}$
extracted as the maximal current in the system. 
($b$) The energy spectra close to $E=0$.
The inset of ($b$) is for the case of $L_y=7$. 
TR symmetry is spontaneously broken between the two dashed red 
lines as evidenced by $J_{max}\neq 0$.
Please note that: 
At large values of $B$, the edge current vanishes which is
an artifact due to the finite length of $L_x$. 
The decaying lengths of edge Majorana modes are at the order
of the superconducting coherence length which is long due to 
the suppression of the pairing gap.
As a result, Majorana modes on opposite edges can hybridize and
are gapped out without breaking TR symmetry. 
\label{fig:L8phasediagram}}
\end{figure}

A natural question is under what conditions TR symmetry is 
spontaneously broken near edges.
We have carried out extensive numerical studies and results 
of $\mu=-2$ are plotted in Fig.~\ref{fig:L8phasediagram} ($a$) 
and ($b$). 
TR symmetry is always broken in the topological gapped phase
such that Majorana edge fermions are pushed to midgap energies,
while TR symmetry remains unbroken in the trivial gapped phase.
The latter is easy to understand because there are no Majorana fermions
to begin with. 
If the bulk is in the gapless phase (shaded area in Fig. 
\ref{fig:L8phasediagram}), the situation is more complicated.
TR symmetry breaking solutions are found in most part of the gapless phase. 
In this regime, $|r|<1$ and thus the number of stable 
Majorana modes on one edge is less than the number of chains $L_y$. 
These modes are associated with the same value of 
$W_{k_y}$, and thus TR symmetry breaking is needed to gap out these edge modes.
There exists a small region inside the gapless phase in which TR 
symmetry is unbroken in  Fig.~\ref{fig:L8phasediagram} ($a$), 
which is largely due to the finite value of $L_y$.
We have tested that as increasing $L_y$ the TR
breaking regime is extended, and thus we expect that it will cover the
entire gapless phase in the thermodynamic limit. 
On the other hand, for the case of the gapless phase with $\mu=0$
in which $r=0$ for even values of $L_y$, our calculations show 
that all the Majorana modes are gapped out without developing 
currents.
Instead a bond-wave order appears at the wavevector 
of $k_y=\pi$ along the edge, which is consistent with 
the fact that TR invariant perturbations can destroy 
Majorana zero energy modes at $r=0$.
In general, we expect that TR symmetry is spontaneously
broken in the case of $r\neq 0$ in the thermodynamic limit.

However, not all Majorana edge modes have to be gapped out
in the topological gapped phase.
As shown in Fig. \ref{fig:L8phasediagram} ($b$),  for the case of $L_y=8$, 
all the edge modes become gapped due to TR symmetry breaking, 
whereas for $L_y=7$, one Majorana mode survives at zero energy.
\footnote{The remaining Majorana mode is localized along $x$-direction, but delocalized along $y$-direction. Similar result can be found in Ref.~\onlinecite{Mizushima2013}.}
The reason is that breaking TR brings the system from class BDI to class 
D \cite{Schnyder2008}, and the latter is characterized by 
a $\mathcal{Z}_2$ index. 
Physically it is because (in the infinite chain length limit) 
only the Majorana modes on the same edge can be paired and gapped out, 
thus beginning with $L_y$ Majorana fermions per edge, for odd $L_y$, one
of them will always remain unpaired.
In short, if TR is \emph{spontaneously} broken, only 
$L_y \textsf{ mod } 2$ Majorana fermion per edge will persist
at zero energy.


\heading{Discussion}%
Before closing, a few remarks are in order. (1) The
  phenomenon of spontaneous TR symmetry breaking in topological
  superconductors has previously been found in a spinless $p$-wave
  superconductor in Ref.~\onlinecite{Li2013}. Our work extends this
  observation in three ways: (a) Our results confirm that spontaneous
  TR breaking also occurs in a different setup with SO
  coupling and s-wave pairing, which is more relevant to experiments. 
  (b) Our model hosts a gapless phase,
  wherein spontaneous TR breaking may also occur. (c) We also found a
  parameter regime where Majorana modes with opposite winding numbers
  can coexist. This provides another route to gap out the Majorana
  modes without invoking TR breaking. (2) In this work, we only
  considered SO coupling in the $x$ direction, which can be exactly
  simulated in cold atom systems. However, in solid state physics,
  both Rashba and Dresselhaus SO couplings will involve SO coupling
  along the $y$ direction as well (unless Rashba and Dresselhaus are
  of equal strength, in which case SO coupling along $y$ will vanish).
  This will break TR symmetry (as defined in Eq.~\ref{eq:pseudoTR},
  which is not the usual physical TR symmetry)
  and bring the system from class BDI to
  D. In the presence of a $y$-direction SO coupling term
  ($\sim\sin(k_y)\sigma_2\tau_3$), the Majorana flat bands will 
  develop dispersion, either connecting upper and lower bulk bands
  or forming isolated mid-gap states 
  which may cross zero at $k_y=0$ or $\pi$, consistent with a
  $\mathcal{Z}_2$ description.\cite{Qu2013,Seroussi2014} (3) Disorders
  such as spatial variations of chemical potential
  ($\sim\sigma_0\tau_3$) and Cooper pairing amplitude
  ($\sim\sigma_2\tau_2$) can be added without changing any of our
  conclusions (provided the disorder is not strong enough to close the
  bulk gap). This is because these two terms are invariant under both
  particle-hole($\Xi$) and TR($\Theta$) symmetries, hence the system
  still belongs to the BDI class. (4) Finally, although we modeled the
  constituent nanowires each as a 1D lattice, switching to a continuum
  formulation in the chain direction should not affect the formation
  of edge Majorana modes (that is, before they couple and gap out).
  \cite{Sau2010,Oreg2010} Thus we expect
  the edge physics obtained here 
  to be insensitive
  to how the bulk of the chains is formulated in terms of continuum
  \emph{vs.~}lattice.

\heading{Summary}
We have studied quantum wire arrays with SO coupling and $s$-wave 
superconductivity in an external Zeeman field.
The relation between edge Majorana zero modes and the bulk 
band structure is investigated in both topologically nontrivial gapped phase
and the gapless phase.
The coupling between Majorana modes and superfluid phases leads to 
spontaneous TR symmetry breaking.
Our results have several experimental bearings. 
For proximity effect induced superconductivity, the number of edge 
Majorana fermions in the gapless phase can be tuned by the
Zeeman field from zero all the way up to the number of chains. 
This could be detected as a prominent change in the height of zero 
bias peaks in tunneling spectroscopy experiments. 
For the intrinsic superconductivity, edge supercurrent loops resulting 
from spontaneous TR breaking will induce small magnetic moments, which 
can be detected using magnetically sensitive experiments such as 
nuclear magnetic resonance or neutron scattering. 
The fluctuation in the number of persisting Majorana mode between 
$1$ and $0$, in the TR-broken topological gapped phase, may also 
show up in tunneling spectroscopy.

\heading{Acknowledgments}
We thank Yi Li for early collaborations, Hui Hu and D.~P.~Arovas for 
helpful discussions, and D.~P.~Arovas for comments after reading a 
draft of this paper. 
DW and CW are supported by the NSF DMR-1105945 and AFOSR FA9550-11-1-
0067(YIP). 
ZSH is supported by NSF through grant DMR-1007028. 
CW acknowledges the support from the NSF of China 
under Grant No. 11328403. 

{\it Note added} Upon the completion of this work, 
we became aware of the nice paper on a similar topic
\cite{Seroussi2014},
and after this work is posted, we noticed another related 
work \cite{wakatsuki2014}.

\bibliography{majorana}

\end{document}